\documentclass{PoS}
\usepackage{wrapfig}
\usepackage{amsmath}
\newif\ifstartedinmathmode
\DeclareRobustCommand{\gotomathmode}[1]{\relax\ifmmode\startedinmathmodetrue\else\startedinmathmodefalse\fi
\ifstartedinmathmode #1 \else $#1$\fi}

\newcommand{\MeV}{\mathrm{MeV}}
\newcommand{\GeV}{\mathrm{GeV}}
\newcommand{\D}{\mathrm{d}}
\newcommand{\Em}{E\hspace*{-6pt}/}
\newcommand{\ee}[2]{\gotomathmode{#1\cdot10^{#2}}}

\title{Fully differential NLO predictions for rare and radiative lepton decays}

\ShortTitle{Rare and radiative lepton decays at NLO}
\PREPRINT{PSI-PR-17-20}
\author{
  \speaker{Y. Ulrich}\\
  Paul Scherrer Institut,\\
  CH-5232 Villigen PSI, Switzerland \\

  Physik-Institut, Universit\"at Z\"urich, \\
  Winterthurerstrasse 190,\\
  CH-8057 Z\"urich, Switzerland\\

  E-mail: \email{yannick.ulrich@psi.ch}
}

\abstract{

We present a general purpose, parton-level Monte Carlo program for the
calculation of the radiative ($L\to\nu\bar\nu l+\gamma$) and rare
($L\to\nu\bar\nu l'+ l^+l^-$) muon and tau decays at NLO in the
effective Fermi theory. In the case of muon, these processes are
irreducible Standard Model backgrounds to searches for lepton flavour
violation at the PSI experiments MEG and Mu3e as they become
indistinguishable from the corresponding signals when the neutrinos
carry little energy.

Furthermore, we argue that fully differential NLO corrections are very
important for the analysis of measurements aiming at the percent level
or better. This is especally true if very stringent phase-space cuts
are applied. To illustrate this, we use a tension between
\textsc{BaBar}'s recent measurement of the radiative tau decay and the
Standard Model prediction as an example of such an analysis. Finally,
we present the branching ratios of the rare tau decay
$\tau\to\nu\bar\nu l' l^+l^-$ at NLO.

We generally find that QED corrections of $\mathcal{O}(10\%)$ are very
well possible.

}
\FullConference{The 19th International Workshop on Neutrinos from Accelerators-NUFACT2017\\
		25-30 September, 2017\\
		Uppsala University, Uppsala, Sweden}

\begin{document}

\section{Introduction}
Muon decays have a long and rich history in particle physics and will
no doubt continue to play an important rule. Their study, both
experimentally and theoretically, helped to precisely determine
standard model quantities like the Fermi constant $G_F$ and severely
constrain lepton flavour violating (LFV) new
physics~\cite{Kuno:1999jp}. While radiative corrections to the Michel
decay are available for many decades, $\mathrm{NLO}$ results for the
radiative $\mu\to\nu\bar\nu e\gamma$ and rare $\mu\to\nu\bar\nu
ee^+e^-$ decay have only been published in the last two
years~\cite{Fael:2015gua,Fael:2016yle,Pruna:2016spf,Pruna:2017upz}.

In the case of $\tau$ decays, precision measurements at the percent
level are available. Especially for the radiative $\tau$ decay there
has been a discrepancy at the level of several $\sigma$.

Here we will review the rare and radiative lepton decays in their role as
SM background processes to the searches for LFV by the experiments
MEG~\cite{Adam:2013gfn} and
Mu3e~\cite{Berger:2014vba,Perrevoort:2016nuv} and further stress the
importance of fully differential calculations.

The framework of our calculation is the QED Lagrangian augmented with
the Fierz rearranged effective Fermi interaction
\begin{align}
\mathcal{L} &
= \mathcal{L}_{\text{QED}} + \frac{4\, G_F}{\sqrt2} 
\left( \bar\psi_e \gamma^\mu P_L \psi_\mu \right)  
\left( \bar\psi_{\nu_\mu} \gamma^\mu P_L \psi_{\nu_e} \right) + \text{h.c.}\, ,
\end{align}
where $P_L=(1-\gamma_5)/2$ is the usual left-handed projector. The QED
part contains the necessary lepton fields, i.e. $\psi_\mu$ and
$\psi_e$. Even though this is an effective field-theoretical
description of the underlying electroweak interaction, it has been
shown by Berman and Sirlin~\cite{BERMAN196220} that QED corrections
are finite at all orders and that $G_F$ does not need to be
renormalised. Unless noted otherwise, we keep the electron mass
$m_e\neq0$. 

We compute the one-loop diagrams with a modified version of
\text{GoSam}~\cite{cullen:2014yla,Mastrolia:2012bu,Peraro:2014cba} as
well as \texttt{FeynArts}~\cite{Hahn:2000kx},
\texttt{FormCalc}~\cite{Nejad:2013ina} and
\texttt{LoopTools}~\cite{Hahn:1998yk}.  Infrared singularities in the
real corrections are dealt with using FKS
subtraction~\cite{Frixione1995Three-jet,Frederix2009Automation}.
Finally, we integrate the phase space using
\texttt{VEGAS}~\cite{Lepage:1980jk}.  This allows us to create
arbitrary distributions involving arbitrary cuts. Further details
regarding the implementation are available in
\cite{Pruna:2016spf,Pruna:2017upz}.

\section{Radiative tau decay $\tau\to \nu\bar\nu e\gamma$}
Aside from being an important background to searches for $\mu\to
e\gamma$, the radiative lepton decays $L\to\nu\bar\nu l+\gamma$ can be
used as an example to illustrate why fully differential Monte Carlo
programs are important. This can be seen best in the radiative tau
decay $\tau\to\nu\bar\nu e+\gamma$ which was measured by
\textsc{BaBar}~\cite{Oberhof2015Measurement}. As observed by Fael,
Mercolli and Passera~\cite{Fael:2015gua}, the measured value and the
$\rm NLO$ branching ratio disagree at around $3.5\sigma$.

Experimentally, the following happens: a pair of tau leptons is created
in $e^+e^-$ collisions at $\sqrt{s} = M_{\Upsilon(4S)} =
10.58~\GeV$.  One of the taus is used for tagging, while the signal is
measured with the other tau. Very stringent cuts are imposed on the
signal tau in the centre-of-mass frame of the $e^+e^-$ pair to reduce
background events
\begin{align}
\cos\theta^*_{e\gamma} &\ge 0.97, &
0.22~\GeV &\le E^*_{\gamma} \le 2.0~\GeV, &
M_{e\gamma} &\ge 0.14~\GeV\, .
\label{eq:cuts}
\end{align}
By implementing these cuts in our Monte Carlo program, we will see,
that the NLO corrections will have an important effect when `undoing'
the cuts, i.e. when extracting exclusive branching ratio\footnote{We
always normalise the branching ration to the measured life time and
not its LO order result} $\mathcal{B}^{\mathrm{excl}}$ (with only the
cut $E_{\gamma}\ge 10~\MeV$ in the tau rest frame).

To account for this effect we use the following simplified scheme:
Assume that $N_{\text{obs}}$ events passed the cuts~(\ref{eq:cuts}).
To transform $N_\text{obs}$ into a branching fraction, it is multiplied
by a factor $\epsilon_{\text{LO}}^{\text{exp}}$ 
\begin{align}
\mathcal{B}^{\mathrm{LO}}_{\text{exp}} 
    = \epsilon_{\mathrm{LO}}^{\text{exp}} \cdot N_{\text{obs}}
    = \epsilon_{\text{det}}\cdot \epsilon_{\mathrm{LO}} \cdot N_{\text{obs}}
    = \ee{1.834(1)}{-2}\,.
\end{align}
Detector effects are assumed to be described by
$\epsilon_{\text{det}}$ while $\epsilon_{\mathrm{LO}}$ is a purely
theoretical factor that converts the fiducial branching ratio to the
desired value. The computation of this factor can be done both at $\rm
LO$ and $\rm NLO$
\begin{align}
\epsilon_{\mathrm{LO}} =
\frac{\Gamma_{\mathrm{LO}}^{10\,\MeV}}
  {\Gamma_{\mathrm{LO}}^{\text{fiducial}}} \bigg|_{\rm theory} =
  48.55(1) 
\,,\qquad
\epsilon_{\mathrm{NLO}} =
\frac{\Gamma_{\mathrm{NLO}}^{10\,\MeV}}
  {\Gamma_{\mathrm{NLO}}^{\text{fiducial}}} \bigg|_{\rm theory} =
  44.80(1)\,,
\end{align}
where $\Gamma^{10\,\MeV}$ ($\Gamma^{\text{fiducial}}$) refers to the
$10\,\MeV$ (fiducial) cut. We can calculate the value of
$\mathcal{B}_{\text{exp}}$ applying an $\rm NLO$ Monte Carlo
\begin{align}
\mathcal{B}^{\mathrm{NLO}}_{\text{exp}}
=
  \frac{\epsilon_{\mathrm{NLO}}}{\epsilon_{\mathrm{LO}}} \cdot
  \mathcal{B}^{\mathrm{LO}}_{\text{exp}}
=
  \epsilon' \cdot \mathcal{B}^{\mathrm{LO}}_{\text{exp}} =
  \ee{1.704(50)}{-2}\,,
\end{align}
with $\epsilon'=0.923(1)$. This reduces the discrepancy from
$3.5\sigma$ to $1.2\sigma$. 

Obviously, this is only a very simplistic and by far not complete
simulation of the full analysis. Hence, we do not claim that this is
the conclusive resolution of the discrepancy. However, we do point out
that even in QED a proper inclusion of $\rm NLO$ effects is mandatory
for a precision around the percent level. Further details of this
analysis can be found in~\cite{Pruna:2017upz}. In light of recent
$B$-anomalies it is worth noting, that QED corrections can be large and
should be included whenever possible.

\section{Radiative muon decay $\mu\to\nu\bar\nu e\gamma$}
Let us first look at the PiBeta experiment~\cite{Pocanic:2014mya},
which used
\begin{align*}
E_\gamma\ge10\,\MeV
\quad\text{and}\quad
\sphericalangle(\vec p_\gamma, \vec p_e)\ > \ 30^\circ\,.
\end{align*}
While the PiBeta measurement offers sub-percent precision, it was
noted before~\cite{Pruna:2017upz,Bernetau} that there is a significant
disagreement between the quoted value for the standard model and our
calculation. However, we can reproduce the quoted standard model value
if we redo our calculation with vanishing electron mass $m_e=0$ and
the inclusion of a factor to account for conversion into an $e^+e^-$
pair~\cite{Eckstein1959297}. Using the approach explained above, we
can convert the PiBeta measurement of $\mathcal{B}_{\pi\beta} =
\ee{4.365(42)}{-3}$ into
\begin{align*}
  \mathcal{B}_{\pi\beta}^* = \epsilon_{\pi\beta}'
  \mathcal{B}_{\pi\beta} = \ee{4.18(4)}{-3}
\end{align*}
which is compatible with the theoretical prediction of
$\mathcal{B}=\ee{(4.26-0.04_{\rm NLO})}{-3}$.

We can use these considerations to compare and contrast all major
measurements of the radiative muon decay that are currently listed in
the PDG. This is shown in Table~\ref{comparison}.

\begin{figure}
\centering
\begin{tabular}{c||c|c|l}
Experiment                                              &
   $\epsilon     $          &
   $          \mathcal{B}_{\text{meas.}}$               &
   $10^2\cdot \epsilon\cdot \mathcal{B}_{\text{meas.}}$
\\\hline
MEG~\cite{Adam:2013gfn}      & \ee{2.2}5 &\ee{6.0(5)}{-8} & 1.3(1)  \\
PiBeta~\cite{Pocanic:2014mya}& 2.9       &\ee{4.18(4)}{-3}& 1.27(1) \\
\cite{Crittenden:1959hm}     & 1.0       &\ee{1.4(4)}{-2} & 1.4(4)
\\\hline
Average & \multicolumn{2}{|c|}{}                          & 1.27(1)
\end{tabular}
\renewcommand{\figurename}{Table}

\label{comparison}
\caption{Various measurements of the radiative muon decay related the
canonical configuration with $E_\gamma\ge10\,\MeV$ using the kinematic
acceptance $\epsilon =\mathcal{B}^{\text{th}}(10\,\MeV)/
\mathcal{B}^{\text{th}}(\text{fiducial})$}

\end{figure}

\begin{figure}
\centering
\setlength{\unitlength}{1cm}
\scalebox{0.8}{\begin{picture}(14,10)
\put( 0.5, 0.5){\includegraphics[width=13cm]{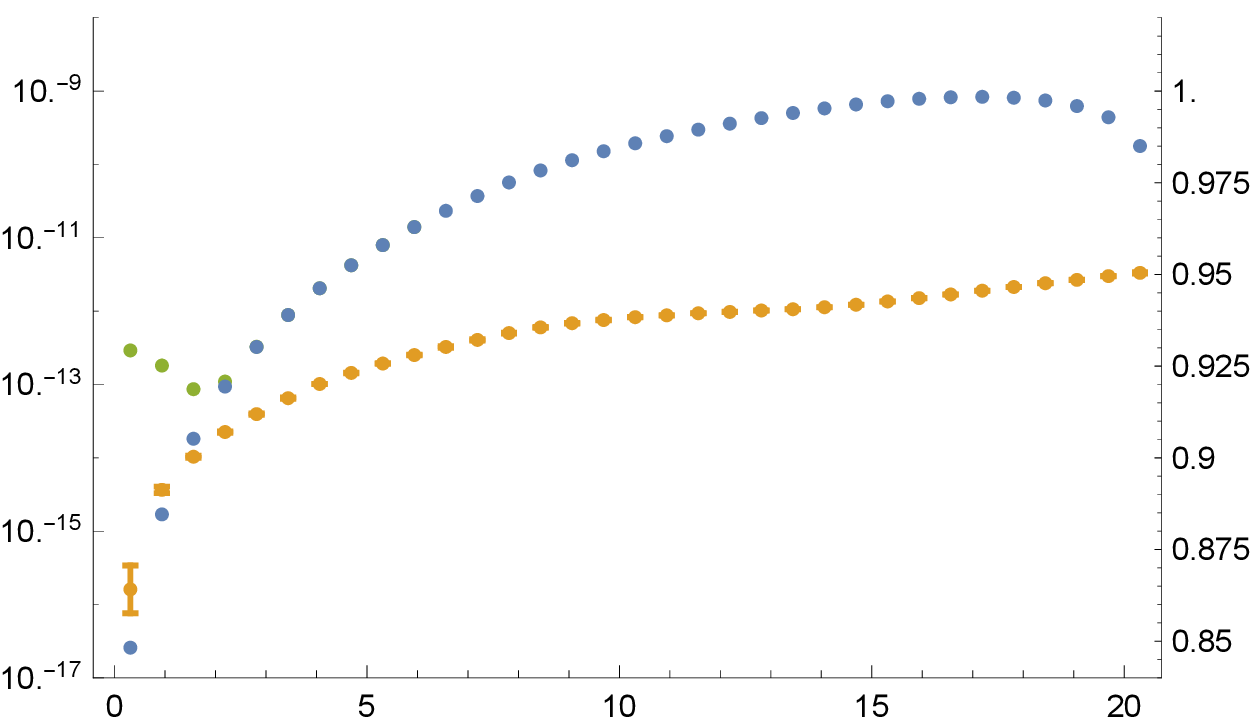}}
\put( 6.0, 0.0){$\Em/\MeV$}
\put( 0.0, 4.0){\rotatebox{90}{$\D\mathcal{B}/\D \Em$}}
\put(13.7, 5.0){\rotatebox{-90}{$K$}}
\end{picture}}
\caption{The invisible energy spectrum for MEG in blue at {\rm NLO},
the $K$-factor as the ratio between {\rm NLO} and {\rm LO} in orange
and in green a mock-up of how LFV would look like at the current limit
$\mathcal{B}^{\text{NP}}\simeq \ee{4.2}{-13}$ accounting for a
1.7\% energy resolution.}
\label{meg}
\end{figure}

For MEG, which uses $85\%$ polarised muons, we define the $\vec z$
axis along the muon polarisation, i.e.  $\vec P_\mu = -0.85\ \vec z$.
We now can model the cuts used by the MEG experiment as
\begin{subequations}
\label{cut:mue}
\begin{align}
E_\gamma > 40\,\MeV\,,\qquad
&E_e     > 45\,\MeV
               \label{eq:cuts:energy}
\,,\\
|\cos\sphericalangle(\vec p_\gamma, \vec z)|
 \equiv|\cos\theta_\gamma| < 0.35\,,\qquad
&      |\phi_\gamma|       > \frac{2\pi}3 
               \label{eq:cuts:geom_photon}
\,,\\
|\cos\sphericalangle(\vec p_e, \vec z)|
 \equiv|\cos\theta_e     | < 0.5\,,\qquad
&      |\phi_e           | < \frac{\pi}3
               \label{eq:cuts:geom_electron}\,.
\end{align}
\end{subequations}
Further, we require that exactly one photon with an energy larger than the
detector threshold of roughly $2\,\MeV$ hits the detector. Therefore,
we reject a second photon that hits the detector if its energy is
larger than $2\,\MeV$, i.e.
\begin{align}
E_{\gamma_2} < \begin{cases}
2\,\MeV & \text{if \eqref{eq:cuts:geom_photon} is satisfied}\\
\infty & \text{otherwise}
\end{cases}
\tag{\theequation d} \,.\label{eq:cuts:soft}
\end{align}
This allows us to calculate for example the distribution w.r.t. the
invisible energy $\Em = m_\mu - E_e - E_\gamma$ as shown in
Figure~\ref{meg}. The corrections are fairly large
$\mathcal{O}(5-10\%)$ but always negative. This means that a
background study that only used a LO calculation will actually produce
a more conservative approximation.

\section{Rare muon decay $\mu\to\nu\bar\nu ee^+e^-$}
\begin{figure}
\centering
\setlength{\unitlength}{1cm}
\scalebox{0.8}{\begin{picture}(14,10)
\put( 0.5,0.5){\includegraphics[width=13cm]{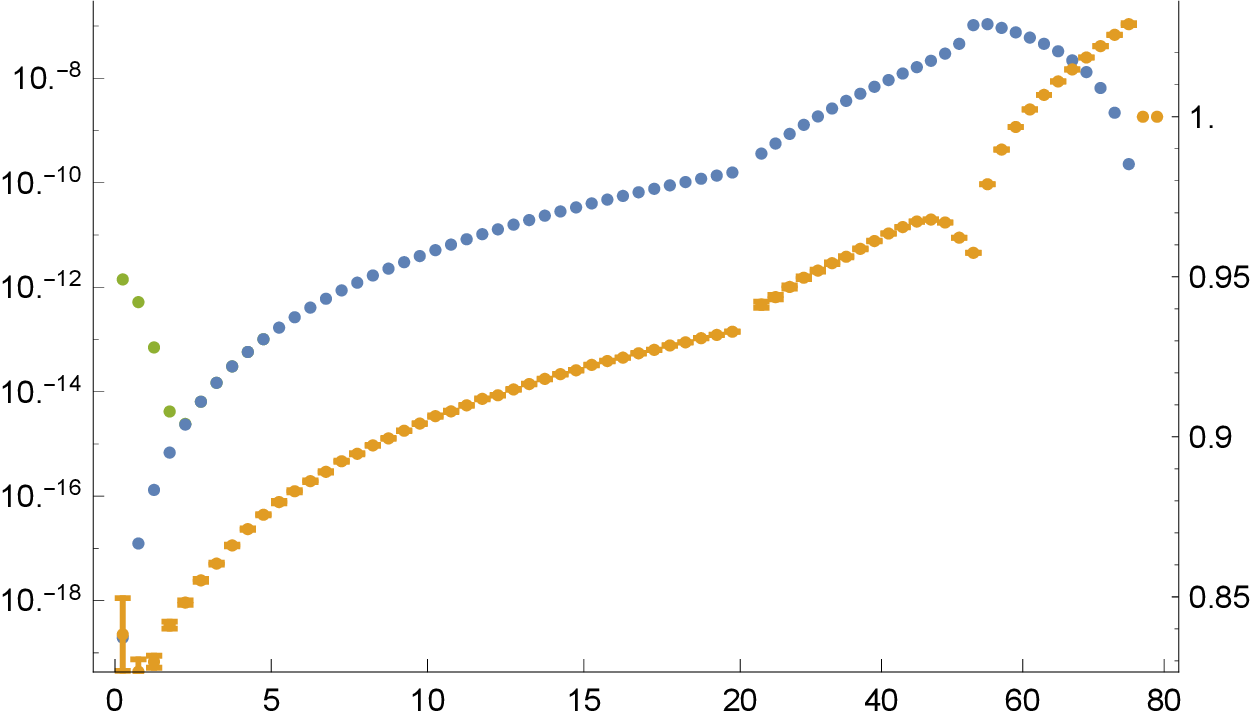}}
\put( 6.0, 0.0){$\Em/\MeV$}
\put( 0.0, 4.0){\rotatebox{90}{$\D\mathcal{B}/\D \Em$}}
\put(13.7, 5.0){\rotatebox{-90}{$K$}}
\end{picture}}
\caption{The invisible energy spectrum for Mu3e in blue at {\rm NLO}, the
$K$-factor as the ratio between {\rm NLO} and {\rm LO} in orange and
in green a mock-up of how LFV would look like at the current limit
$\mathcal{B}^{\text{NP}}\simeq 10^{-12}$ accounting for a
$0.5\,\MeV$ energy resolution.}
\label{mu3e}
\end{figure}
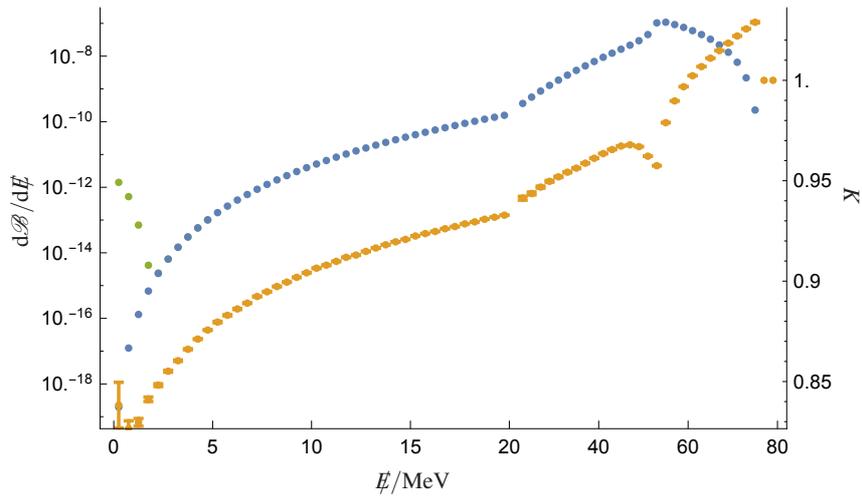

For the rare decay we use
\begin{align}
E_{e^\pm} > 10\,\MeV
\quad\text{and}\quad
|\cos\sphericalangle(\vec p_{e^\pm}, \vec e_z)| < 0.8
\end{align}
to roughly approximate the Mu3e detector~\cite{Perrevoort:2016nuv}. We
can again create the invisible energy distribution as shown in
Figure~\ref{mu3e}. The corrections are again very large for QED but
still negative in the region of any potential signal.

While the observation of $\mu\to eee$ would be a clear indication of
physics beyond the standard model, we would like to propose a
different way to use a precise measurement of the rare muon decay to
search for light new mediators: the Lorentz structure of a new
mediator could be encoded in angular distributions of the $e^+e^-$
pair. This is especially promising because theoretical uncertainties
on these observables are very small as they only receive small
corrections at the percent level. Furthermore, these corrections are
very flat and barely distort the distribution's shape. Therefore, we
can expect that the $\rm NNLO$ corrections to the shape would be well
below the per-mille level.

\section{Rare tau decay $\tau\to\nu\bar\nu l'l^+l^-$}
The rare tau decay $\tau\to\nu\bar\nu l'll$ has a richer phenomenology
than the muon decay because more decay channels are possible. In fact,
a plethora of tree-level calculations have been published using a
variety of
techniques~\cite{Flores-Tlalpa:2015vga,Arroyo-Urena:2017ihp,Dicus:1994dt}.
On the other hand, experimental data is scarce. Nevertheless, we have
used our program to calculate all four rare leptonic tau decay
channels at NLO.  Our results are summarised in Table~\ref{raretau}
and compared with \cite{Arroyo-Urena:2017ihp}. While our result
includes all mass-effects it is not clear how
reference~\cite{Arroyo-Urena:2017ihp} treated the light lepton's mass
which cannot be set to zero because that would give rise to IR
singularities. When computing the radiative corrections we have all
three lepton-flavours active but neglect hardronic corrections because
a rough estimate shows that they do not change the overall
picture~\cite{Pruna:2016spf}.

\begin{figure}
\centering
\begin{tabular}{c|c|c|c}
 &    \multicolumn{2}{c|}{This work}  & \cite{Arroyo-Urena:2017ihp} \\
 & LO & $\delta\mathcal{B}/\mathcal{B}$          & \\\hline

$10^5\mathcal{B}(\tau\to\nu\bar\nu eee       )$& 4.2489(1) &  -0.094\%& 4.22(2)  \\
$10^7\mathcal{B}(\tau\to\nu\bar\nu\mu ee     )$& 1.9879(2) &   1.1  \%& 1.987(3) \\
$10^5\mathcal{B}(\tau\to\nu\bar\nu\mu \mu e  )$& 1.2513(2) &   1.2  \%& 1.246(2) \\
$10^7\mathcal{B}(\tau\to\nu\bar\nu\mu \mu \mu)$& 1.1838(1) &   1.9  \%& 1.184(1) 

\end{tabular}
\renewcommand{\figurename}{Table}

\caption{The branching ratio of the rare tau decays computed at LO (by
us and \cite{Arroyo-Urena:2017ihp}) and NLO. We neglect hadronic
corrections.}
\label{raretau}

\end{figure}

\section{Conclusion}
We have reviewed the $\rm NLO$ QED predictions for radiative $L\to
l\nu\bar\nu+\gamma$ and rare $L\to l\nu\bar\nu+l^+l^-$ lepton
decays. In particular, we emphasise the importance of radiative
corrections in unfolding fiducial acceptance when comparing
experimental values with the PDG.

Furthermore, our program is able to calculate arbitrary differential
distributions. For the experiments MEG and Mu3e, the corrections to
these distributions is usually at the percent level. But especially in
the critical regions of phase space they can easily reach
$\mathcal{O}(10\%)$.

\acknowledgments{
The author is grateful to Matteo Fael and Massimo Passera for
comparing and discussing our calculations. Furthermore, we are
indebted to Angela~Papa, Giada~Rutar, Ann-Kathrin~Perrevoort,
Alberto~Lusiani and Dinko~Pocanic for their explaination of the
experimental details underlying MEG, Mu3e, \textsc{BaBar} and Pibeta.
The author is supported by he Swiss National Science Foundation (SNF)
under contract 200021\_163466.
}

\bibliographystyle{JHEP}
\bibliography{../../muon_ref.bib}{}

\end{document}